\documentclass[aps,prb,twocolumn,groupedaddress,floatfix,showpacs]{revtex4-2}
\usepackage[T1]{fontenc}
\usepackage{graphicx} 
\usepackage{amsmath}
\usepackage{color}

\begin{document}

%\preprint{}

\title{Electro-optical properties of excitons in Cu$_2$O
quantum wells: II continuum states}


\author{David Ziemkiewicz}
\email{david.ziemkiewicz@utp.edu.pl}
\author{Gerard Czajkowski}
\author{Karol Karpi\'{n}ski}
\author{Sylwia
Zieli\'{n}ska-Raczy\'{n}ska}
%\thanks{}
 \affiliation{Institute of
Mathematics and Physics, UTP University of Science and Technology,
\\Aleje Prof. S. Kaliskiego 7, 85-789 Bydgoszcz, Poland.}


\date{\today}

\definecolor{green}{rgb}{0,0.8,0}

\begin{abstract}
We present  theoretically calculated the
optical functions for Cu$_2$O quantum well (QW) with Rydberg
excitons  in an external homogeneous electric field parallel to the QW planes for the energy region above the gap, suitable to observe the Franz-Keldysh (FK) oscillations.
We quantitatively describe the amplitudes and periodicity of FK modulations and the influence of both, Rydberg excitons and confinement effect, on this phenomenon.

% Two configurations of an external electric field
%perpendicular and paralell to the QW planes, and the field parallel  are considered in the energetic region of
%\textit{p}-excitons and for discrete.% and continuum states.
%With the help of
%the real density matrix approach and an effective e-h potential,
%which enable to derive the analytical expressions for the QW
%electro-optical functions are calculated  for the case of the excitation energy below  the gap energy.
%energy.
%For both configurations, all three
%field regimes: field perpendicular, field parallel and the
%excitation energy smaller than the gap energy (discrete states),
%nd the field parallel, the excitation energy larger than the gap
%energy (continuum states) are considered and treated separately.

\end{abstract}
 \maketitle
\section{Introduction}
In recent years there has been much interest in the optical properties of Rydberg excitons (RE) in bulk semiconductors in the context of their unusual capabilities while interacting with external fields \cite{Kazimierczuk,AssmannBayer_2020}, with particular applications in quantum information processing \cite{Khazali,Walther}.
  Later, the studies were extended to the electro-optic properties of bulk Cu$_2$O
crystals in the energy region above the fundamental gap, where the
oscillations appear \cite{SZR FK, pss} which is often referred as the Franz-Keldysh effect \cite{Franz, Keldysh}.
%These oscillations have  emerged from the leaking wave functions into the %the band gap, which manifestes as a weak absorption tail. The appearance %of this structure is 

An externally applied electric field affects the absorption edge via Franz-Keldysh effect, which is the result of wavefunctions "leaking" into the band gap. Due to the electric field the electron and hole wave functions become Airy functions, which are characterized by a "tail"  extending into the classically forbidden band gap. The absorption spectrum is modified: a tail at energies below the band gap and some oscillations above it are observed.
  
This phenomenon  was considered before in low dimensional structures such as quantum wires, nanobelts \cite{Cavalini,Li,Xia3} and carbon nanotubes \cite{Perebeinos}. Miller \textit{et al} \cite{Miller} studied the confined Franz-Keldysh effect in quantum wells with electric field along the direction of carrier confinement. Xia and Spector \cite{Xia,Xia2} presented theoretical studies of Franz-Keldysh effect in the interband optical absorption for various nanostructures: quantum wires, wires and boxes. However, the excitonic effects were neglected in these papers.

The topic of FK effect in low dimensional nanostructures particularly with RE has not been covered yet. In our previous work \cite{DZ} we have studied quantum wells with Rydberg excitons interpose in an external electric field below the gap. In this work we are concentrated on the case in which an electric field is oriented parallel to the direction of the quantum well plane for excitation energies region above the band gap. In such situation gives rise to Franz–Keldysh oscillations analogous to these bulk case. Though in the quantum well physical picture becomes more complicated because there is an interplay between the strong Coulomb interaction of particles forming the exciton, a constraint superimposes by confinement and above the gap excitation.  Moreover, the effect of excitons may influence the optical properties near the band gap \cite{Dow} and the situation might become more complex when one considers conditions superimposed by quantum wells potential barriers \cite{Perebeinos}. Even though excitonic impact is not very important for the position of resonances it is significant for the magnitude of the absorption.
 
Cuprous oxide raises  prospect of a material where Rydberg excitons in quantum wells  can be created \cite{Hamid}. In the future it can be possible to fabricate  low-dimensional structures, which can provide capabilities to create scalable quantum devices. Taking advantage of above the gap oscillations typical for Franz-Keldysh effect might pave the way to designate the flexible electro-modulators.  
The theoretical description of the FK effect in Cu$_2$O  quantum wells with RE requires the continuum states have to be taken into account. With the help of real density matrix approach we will derive the analytical formulas for  electro-optical susceptibility for QW without and with RE and periodicity of modulations. 

The paper is organized as follows.  In Section II we will present the formal theory of electro-absorption in a quantum well  for the energy region above the fundamental gap.  Section III is contains the presentation of illustrative results of Franz-Keldysh oscillations, discussing the effects of multiple confinement states arising from limited QW thickness and the effect of excitonic states on the spectrum above the gap. Finally, in section IV general conclusions are presented.

\section{Theory}
We consider the case of quantum wells when the electric field is
applied parallel to the QW planes (a lateral electric field) and our
aim is to describe  its optical response in the situation where 
electromagnetic wave propagates in the $z$ direction and excited states above the gap.
As it was described in our first paper \cite{DZ}  a Cu$_2$O quantum well  of
thickness $L$, located in the $xy$ plane, with QW surfaces located
at $z=\pm L/2$ is considered. A linearly polarized electromagnetic
wave of the frequency $\omega$ and an electric field vector
\textbf{E} is incident normally on the QW. We will consider
here the situation where the electric field is parallel to the
plane of the well and it is perpendicular to the direction of the
confinement. \noindent

The problem  will be examined using the real density matrix approach (RDMA), as it was done in our preceding paper \cite{DZ}.
There the scheme which enables one to
calculate the electro-optical functions for cuprous oxide QW with
RE was described for the case of excitation energy  smaller
then the gap energy was presented step by step and here  considering the situation of excitation energy above the gap and we  evoke only the main points of the derivation. \noindent

The optical functions will be obtained by solving a system
of integro-differential equations, consisting of the so-called
constitutive equation and the Maxwell equation. The constitutive equation has the form 
\begin{equation}\label{constitutiveHL}
(H_{QW}-\hbar\omega-i{\mit\Gamma})Y=\textbf{ME},
\end{equation}
which for the electric field \textbf{F} applied in the $x$- direction
has the form
\begin{eqnarray}\label{constitutiveEfield}
&&\partial_tY+\frac{i}{\hbar}\biggl[E_g-\frac{\hbar^2}{2m_e}\partial_{z_e}^2-\frac{\hbar^2}{2m_h}\partial_{z_h}^2\nonumber\\
&&-\frac{\hbar^2}{2\mu}\partial_x^2-\frac{\hbar^2}{2\mu}\partial_y^2+eFx+V_{eh}(x,y)+V_e(z_e)\\
&&+V_h(z_h)-i{\mit\Gamma}\biggr]Y=\frac{i}{\hbar}\textbf{ME}.\nonumber\end{eqnarray}
The confinement potentials $V_e,V_h$ are taken in the parabolic
form
\begin{equation}\label{confinement}
V_{\hbox{\tiny
conf}}=\frac{1}{2}m_e\omega_{ez}^2z_e^2+\frac{1}{2}m_h\omega_{hz}^2z_h^2,\end{equation}
 and the e-h interaction
potential in the two-dimensional form
$V_{eh}=-\frac{e^2}{4\pi\epsilon_0\epsilon_b\rho}$ . The above
equation will be solved by the method described elaborately in
Ref.\cite{SZR FK}. Here we will adopted it for the case of quantum wells and for
the considered configuration presenting in detail all main steps of calculations. We separate in the Hamiltonian of
Eq. (\ref{constitutiveEfield}) a "kinetic+electric field" part
$H_{\hbox{\tiny kin+F}}$ and the electron-hole interaction term
$V$, to obtain
\begin{equation}
H_{\hbox{\tiny kin+F}}Y= ME-VY,
\end{equation}
which gives the Lippmann-Schwinger equation, once the Green
function $G$ appropriate to the "kinetic+electric field" is known
\begin{equation}\label{L_S}
Y=GME-GVY.
\end{equation}
This can be solved by the Green function. % derived inRef.\cite{SZR FK}
Considering the electron and hole confinement
states $N_e,N_h$ the Green function will have the form
\begin{eqnarray}\label{eq:green}
&&G=\sum_{N_e,N_h}\psi^{(1D)}_{\alpha_e,N_e}(z_e)\psi^{(1D)}_{\alpha_e,N_e}(Z)(z_e')\psi^{(1D)}_{\alpha_h,N_h}(z_h)
\psi^{(1D)}_{\alpha_h,N_h}(z_h')\nonumber\\
&&\times\frac{2}{\pi}\int\limits_0^\infty dq
\sin\,qy'\sin\,qy\,g_{N_eN_hq}(x,x'),
\end{eqnarray}
with
\begin{eqnarray}\label{gxzeta}
& &g_{N_eN_hq}(x,x')=g_{N_eN_h}^<\;g_{N_eN_hq}^>,\nonumber\\
\nonumber & &g_{N_eN_hq}^<=\frac{\pi}{f^{1/3}}{\rm
Bi}\left[f^{\frac{1}{3}}
\left(x^<+\frac{\kappa_{N_eN_h}^2+q^2}{f}\right)\right]\nonumber\\
&&+ { i}{\rm Ai}\left[f^{\frac{1}{3}}
\left(x^<+\frac{\kappa_{N_eN_h}^2+q^2}{f}\right)\right],\\
\nonumber & &g^>={\rm Ai}\left[f^{\frac{1}{3}}
\left(x^>+\frac{\kappa_{N_eN_h}^2+q^2}{f}\right)\right].\nonumber
\end{eqnarray}
\noindent  $\rm{Ai}(x)$ and $\rm{Bi}(x)$ are the Airy functions (see, for
example, Ref.\cite{Abramowitz}) and
\begin{eqnarray}\label{eq:kappa}
\kappa_{N_eN_h}^2=\frac{2\mu}{\hbar^2}a^{*2}\left(E_g+W_{N_e}+W_{N_h}-\hbar\omega-i{\mit\Gamma}_{N_eN_h}\right).\nonumber\\
\end{eqnarray}
$W_{N_e}$,$W_{N_h}$ are the confinement energies of the electron and the hole.
We introduce the parameter $f$, which defines the relative electric field strength  and the ionization field 
\begin{equation}
f=\frac{F}{F_{\rm I}}, \,\,\,\,\,\,\,\,\,F_{\rm{I}}=\frac{\hbar^2}{2\mu_{\parallel
}ea^{*3}}=\frac{R^*}{a^*e},
\end{equation}
with excitonic Rydberg $R^*$ and corresponding excitonic Bohr radius $a^*$. 
The full set of used parameters is shown in the table \ref{parametervalues1}.
\begin{table}[ht!]
\caption{\small Band parameter values for Cu$_2$O, Rydberg energy
and excitonic radius calculated from effective masses; masses in free electron
mass $m_0$, the ionization field $F_{\rm I}=R^*/(ea^*)$}\label{tab1}
\begin{center}
\begin{tabular}{p{.2\linewidth} p{.2\linewidth} p{.2\linewidth} p{.2\linewidth} p{.2\linewidth}}
\hline
Parameter & Value &Unit&Reference\\
\hline $E_g$ & 2172.08& meV&\cite{Ziemkiewicz_2020}\\
$R^*$&87.78&meV&\cite{Ziemkiewicz_2020}\\
$\Delta_{LT}$&$1.25\times 10^{-3}$&{meV}&\cite{Ziemkiewicz_2020}\\
$\mit\Gamma$&$3.88/(j+1)^3$&meV&\cite{maser}\\
$m_e$ & 1.0& $m_0$&\cite{Ziemkiewicz_2020}\\
$m_h$ &0.7&  $m_0$&\cite{Ziemkiewicz_2020}\\
$M_{\hbox{\tiny tot}}$&1.56&$m_0$&\cite{Ziemkiewicz_2020}\\
$\mu$ & 0.363 &$m_0$&\cite{Ziemkiewicz_2020}\\
$a^*$&1.1& nm&\cite{Ziemkiewicz_2020}\\
$r_0$&0.22& nm&\cite{Ziemkiewicz_2020}\\
$\epsilon_b$&7.5 &&\cite{Ziemkiewicz_2020}\\
 $F_{\rm I}$&1.02$\times\,10^{3}$&kV/cm\\
\hline
\end{tabular} \label{parametervalues1}\end{center}
\end{table}
The confinement functions
$\psi^{(1D)}_{N_e}(z_e), \psi^{(1D)}_{N_h}(z_h)$  are the quantum
oscillator eigenfunctions of the form
\begin{eqnarray*}\label{eigenf1doscillator}
&& \psi^{(1D)}_{\alpha,N}(z)=
 \pi^{-1/4}\sqrt{\frac{\alpha}{2^N N!}} H_N(\alpha z)
e^{-\frac{\alpha^2}{2}z^2}, \\
&&\alpha = \sqrt{\frac{m \omega_z}{\hbar}},
\end{eqnarray*}
where $H_N(x)$ being Hermite polynomials $(N=0,1,\ldots)$.
The dipole density $M$ is given by the formula
\begin{equation}\label{eq:dipole}
M=\frac{M_0}{\rho_0^2}\delta(x)ye^{-\frac{y^2}{2\rho_0^2}}\delta(z_e-z_h).
\end{equation}
The Lippmann-Schwinger equation (\ref{L_S}) is an integral
equation, which can be solved in several  ways. We choose
the method of a one-parameter probe function and present in detail the way of conducting these intricate calculations.\noindent 

To make sure about the correctness of our results, we will use independently two alternative approximations and two expressions for
the probe functions and two forms for the electron-hole attraction potential. Here we note that for excitation energies above the gap and with applied electric field, the carriers move freely and the attraction potential has a small impact on optical properties. Thus, as will be shown later, both approaches provide very similar results.

In the  first approximation  probe function has the form
\begin{equation}\label{probe1}
Y=Y_0y\,e^{-\kappa_{00}\sqrt{x^2+y^2}}\psi^{(1D)}_{\alpha_e,0}(z_e)\psi^{(1D)}_{\alpha_h,0}(z_h),
\end{equation}
with $\kappa_{00}$ given by Eq. (\ref{eq:kappa}) and unknown parameter $Y_0$, which will be obtained from Eq. (\ref{L_S}). The e-h potential, used in Eq. (\ref{L_S}) together
with the function from Eq.(\ref{probe1}), can be written in scaled variables the following in form 
\begin{equation}
V=-\frac{2}{\sqrt{x^2+y^2}}.
\end{equation}
 The function $Y$ calculated in such a way is
then inserted into the expression for the
 total polarization of the medium %which, in the considered
 %configuration, has the form
\begin{equation}\label{Polar}
P(Z)=2 \hbox{Re}\int d^3{r}\,{\bf M}({\bf r}) Y(Z,{\bf r}),
\end{equation}
where $Z$ is the center-of-mass coordinate. Using the long-wave
approximation, we obtain the position-dependent susceptibility
\begin{equation}\label{nonlocalchi}
\chi(Z)=\frac{P(Z)}{\epsilon_0E(Z)}.
\end{equation}  
The optical properties of the whole layer depend on the mean effective susceptibility
\begin{equation}\label{meaneffchi}
\chi=\frac{1}{L}\int\limits_{-L/2}^{L/2}\frac{P(Z)}{\epsilon_0E(Z)}\;dz,
\end{equation}
which, basing on the Eqs. (\ref{eq:green}) and  (\ref{eq:dipole}), can be written in the form
\begin{eqnarray}\label{chi2D_Q}
&&\chi=\frac{2}{\epsilon_0}\int\limits_{-L/2}^{L/2}\frac{1}{Q}M^*GM\,dZ\nonumber\\
&&=\frac{8}{3f^{1/3}}\frac{\epsilon_b\Delta_{LT}}{R^*}e^{4\rho_0}
\left(\frac{a^*}{L}\right)\nonumber\\
&&\times\sum\limits_{N=0}^{N_{\hbox{\tiny{max}}}}\langle\Psi_{NN}\rangle_L
\frac{1}{Q}\int\limits_0^\infty
dq\,q^2\,e^{-\rho_0^2q^2}\times\biggl[{\rm
 Bi}\,\left(\frac{\kappa_{NN}^2+q^2}{f^{2/3}}\right)\,\nonumber\\
 &&\times{\rm
 Ai}\left(\frac{\kappa_{NN}^2+q^2}{f^{2/3}}\right)+i{\rm
 Ai}^2\left(\frac{\kappa_{NN}^2+q^2}{f^{2/3}}\right)\biggr].
\end{eqnarray}
The resonant denominator $Q$ is defined as
\begin{eqnarray}\label{eq:denomin}
&&Q=1-\frac{MGVY}{MY}=\nonumber\\
&&=1-\left[\exp\left(\frac{\kappa_{00}^2\rho_0^2}{4}\right)D_{-3}\left(\kappa_{00}\rho_0\right)\right]^{-1}
\frac{\sqrt{2\pi}}{f^{1/3}}\nonumber\\
&&\times\Biggl\{\int\limits_0^\infty
dq\,\frac{q^2}{\sqrt{q^2+\kappa_{00}^2}}e^{-q^2\rho_0^2/2}{\rm
Ai}\left(\frac{\kappa_{N_eN_h}^2+q^2}{f^{2/3}}\right)\\
&&\times\int\limits_0^\infty dx\, x\, K_1\left(
x\,\sqrt{q^2+\kappa_{00}^2}\right)\biggl[{\rm Bi}
\left(\frac{\kappa_{N_eN_h}^2+q^2}{f^{2/3}}-f^{1/3}x\right)\nonumber\\
&&+ { i}{\rm Ai}
\left(\frac{\kappa_{N_eN_h}^2+q^2}{f^{2/3}}-f^{1/3}x\right)\biggr]\nonumber\\
&&+\int\limits_0^\infty
dq\,\frac{q^2}{\sqrt{q^2+\kappa_{00}^2}}e^{-q^2\rho_0^2/2}\biggl[{\rm
Bi}
\left(\frac{\kappa_{N_eN_h}^2+q^2}{f^{2/3}}\right)\nonumber\\
&&+i{\rm
Ai}\left(\frac{\kappa_{N_eN_h}^2+q^2}{f^{2/3}}\right)\biggr]\nonumber\\
&&\times\,\int\limits_0^\infty dx\, x\, K_1\left(
x\,\sqrt{q^2+\kappa_{00}^2}\right){\rm
Ai}\left(\frac{\kappa_{N_eN_h}^2+q^2}{f^{2/3}}+f^{1/3}x\right)\Biggr\},\nonumber
\end{eqnarray}
where $D_\nu(z)$ is the parabolic cylinder function
\cite{Abramowitz}, and $K_1(z)$ is the modified Bessel function of
the second kind.

Regarding the properties of Airy functions it is known, that for
 negative argument they show oscillatory behaviour. As follows from
 Eq. (\ref{chi2D_Q}), the absorption being the imaginary part of the susceptibility, will show oscillations, known
 as Franz-Keldysh  oscillations (Franz-Keldysh effect). Considering
 the quantum well geometry this effect is also termed
 2-dimensional FK effect (or alternatively lateral FK effect).

To point out the differences between 2- and 3-dimension structures we can compare the absorption of quantum
well and bulk systems \cite{SZR FK}. As a first approximation, one can omit the resonant denominator $Q$; by examining the Eq. (\ref{eq:denomin}) one can see that $Q \approx 1$ is a particularly good approximation in the limit of large $f$ or energy far above $E_g$, which results in large $\kappa_{00}^2$ in the first exponential function of Eq. (\ref{eq:denomin}).
With such assumption, for the 2-dimensional system one obtains
\begin{eqnarray} \label{chi2D}
&&\hbox{Im}\,\chi^{2D}=\chi'\left(\frac{a^*}{L}\right)\sum\limits_{N=0}^{N_{\hbox{\tiny{max}}}}\langle\Psi_{NN}\rangle_L\exp\left(-{\mathcal
E}_Nf^{2/3}\right)f^{2/3}\nonumber\\
&&\times\int\limits_{-{\mathcal E}_N}^\infty du\,\left(u+{\mathcal
E}_N\right)^{1/2}\,e^{-\rho_0^2f^{2/3}u}{\rm
 Ai}^2(u),\nonumber\\
 &&{\mathcal E}_N=\lim\limits_{{\mit\Gamma}\to
 0}\frac{\hbar\omega-E_g-W_{N_e}-W_{N_h}+i{\mit\Gamma}}{\hbar\Theta},\nonumber\\
&& N_e=N_h=N,\qquad \hbar\Theta=R^*f^{2/3},\\
 &&\chi'=\frac{4}{3}\frac{\epsilon_b\Delta_{LT}}{R^*}e^{4\rho_0},\nonumber
\end{eqnarray}
while for the bulk case the susceptibility was described by the following formula \cite{SZR FK}
\begin{eqnarray}
&&\hbox{Im}\,\chi^{3D}=\chi'\exp\left(-{\mathcal E}f^{2/3}\right)\,f\nonumber\\
&&\times\int\limits_{-{\mathcal E}}^\infty du\,\left(u+{\mathcal E}\right)\,e^{-\rho_0^2f^{2/3}u}{\rm
Ai}^2(u),\nonumber\\
&&{\mathcal E}=\lim\limits_{{\mit\Gamma}\to
0}\frac{\hbar\omega-E_g+i{\mit\Gamma}}{\hbar\Theta}.
\end{eqnarray}

The maxima of oscillations for the QW are estimated as
\begin{equation}\label{chi_okres}
E_{Nn}=E_g + W_{ne} + W_{nh} + \left(\frac{3}{3}N\pi\right)^{2/3}\hbar\theta.
\end{equation}
They are similar as these in the bulk case \cite{SZR FK} but modified by the addition of constant energy shift  $W_{ne} + W_{nh}$.
The precise calculation of the resonant term $Q$ as shown in Eq. (\ref{eq:denomin}) can be numerically challenging due to the multiple integrals containing Airy functions, which exhibit oscillatory behaviour and cannot be easily truncated to finite integration limits. Moreover, their arguments diverge for $f\to 0$, which leads to infinite value of $B_i$ and indefinite value of $A_i$. Thus, a more robust approach may be needed for some cases. One can take advantage of the fact that the excitons provide only a small correction to the F-K spectrum, which is relatively insensitive to the form of Coulomb potential. Therefore, one of the options is to use a different electron-hole potential of the form
\begin{equation}
V(x,y)=V_0\,e^{-v^2y^2}\delta(x),
\end{equation} 
with the corresponding function $Y$
\begin{equation} Y=Y_0y\exp\left[-\vert \kappa_{00}^2\vert
(x^2+y^2)\right]\Psi_{00}(z_e,z_h).
\end{equation} 
After substituting them to Eq. (\ref{eq:denomin})) one gets
\begin{eqnarray*}
&&MGVY=(M_0\rho_0)Y_0V_0\left[2(\vert\kappa_{00}^2\vert+v^2)\right]^{-3/2}\frac{\pi}{2}f^{1/3}\\
&&\times\int\limits_{-{\mathcal E}_0}^\infty \sqrt{u+{\mathcal
E}_0}e^{-sf^{2/3}(u+{\mathcal E}_0)}[{\rm Bi}(u)+i{\rm Ai}(u)]{\rm Ai}(u)du,
\end{eqnarray*}
and
\begin{eqnarray}\label{uproszczenie}
&&\frac{MGVY}{MY}=
V_0\left\{\frac{1+2\vert\kappa_{00}^2\vert\rho_0^2}{2(\vert\kappa_{00}^2\vert+v^2)}\right]^{3/2}\sqrt{\frac{\pi}{2}}
f^{1/3}\nonumber\\
&&\times\biggl\{\int\limits_0^{{\mathcal
E}_0}e^{-sf^{2/3}({\mathcal E}_0-u)}\sqrt{{\mathcal E}_0-u}\;[{\rm
Bi}(-u)+i{\rm
Ai}(-u)]{\rm Ai}(-u)du\nonumber\\
&&+\int\limits_{0}^\infty e^{-sf^{2/3}({\mathcal
E}_0+u)}\sqrt{u+{\mathcal E}_0}\;[{\rm Bi}(u)+i{\rm Ai}(u)]{\rm
Ai}(u)du\biggr\},\\
&&s=\frac{\rho_0^2}{2}+\frac{1}{4(\vert\kappa_{00}^2\vert+v^2)},\nonumber\\
&&{\mathcal E}_0=-\frac{\kappa_{00}^2}{f^{2/3}}.\nonumber
\end{eqnarray}
and in the limit of $f\to 0$
\begin{eqnarray*}\label{inny_pot}
&&MGVY_{f\to 0}\approx
(M_0\rho_0)Y_0V_0\frac{\sqrt{2\pi}}
{4\kappa_{00}}\left[\frac{(\vert\kappa_{00}^2\vert+v^2)}{2(\vert\kappa_{00}^2\vert+v^2)\rho_0^2+1}\right]^{3/2},\\
&&\frac{MGVY}{MY}=\frac{(M_0\rho_0)Y_0V_0\frac{\sqrt{2\pi}}{4\kappa_{00}}\left[\frac{(\vert\kappa_{00}^2\vert+v^2)}
{2(\vert\kappa_{00}^2\vert+v^2)\rho_0^2+1}\right]^{3/2}}
{(M_0\rho_0)Y_0\frac{\sqrt{2\pi}}{2}\left(1+2\vert\kappa_{00}^2\vert\rho_0^2\right)^{-3/2}}\\
&&=\frac{V_0}{2\kappa_{00}}\left[\frac{(\vert\kappa_{00}^2\vert+v^2)(1+2\vert\kappa_{00}^2\vert\rho_0^2)}
{1+2(\vert\kappa_{00}^2\vert+v^2)\rho_0^2}\right]^{3/2}.
\end{eqnarray*}
Finally, one arrives at a much simplified expression for $Q$
\begin{equation}\label{eq:denomin2}
Q=1-\frac{MGVY}{MY}=1-\frac{V_0}{2\kappa_{00}}\left[\frac{(\vert\kappa_{00}^2\vert+v^2)(1+2\vert\kappa_{00}^2\vert\rho_0^2)}{1+2(\vert\kappa_{00}^2\vert+v^2)\rho_0^2}\right]^{3/2},
\end{equation}
which is then used in Eq. (\ref{chi2D_Q}). One can see that the Eq. (\ref{eq:denomin2}) is derived in the limit of $f \to 0$, which is the region where the Eq. (\ref{eq:denomin}) fails. In the above equation, the general influence of the excitons on the spectrum is easier to deduce. Overall, $Q \sim 1 - \Delta Q$, with $\Delta Q \sim \kappa_{00}^2 \sim \hbar\omega$. Therefore, the value of susceptibility is slightly larger, with the difference from no exciton spectrum increasing with energy.
\section{Results}
\begin{figure}[ht!]
\centering
a)\includegraphics[width=.8\linewidth]{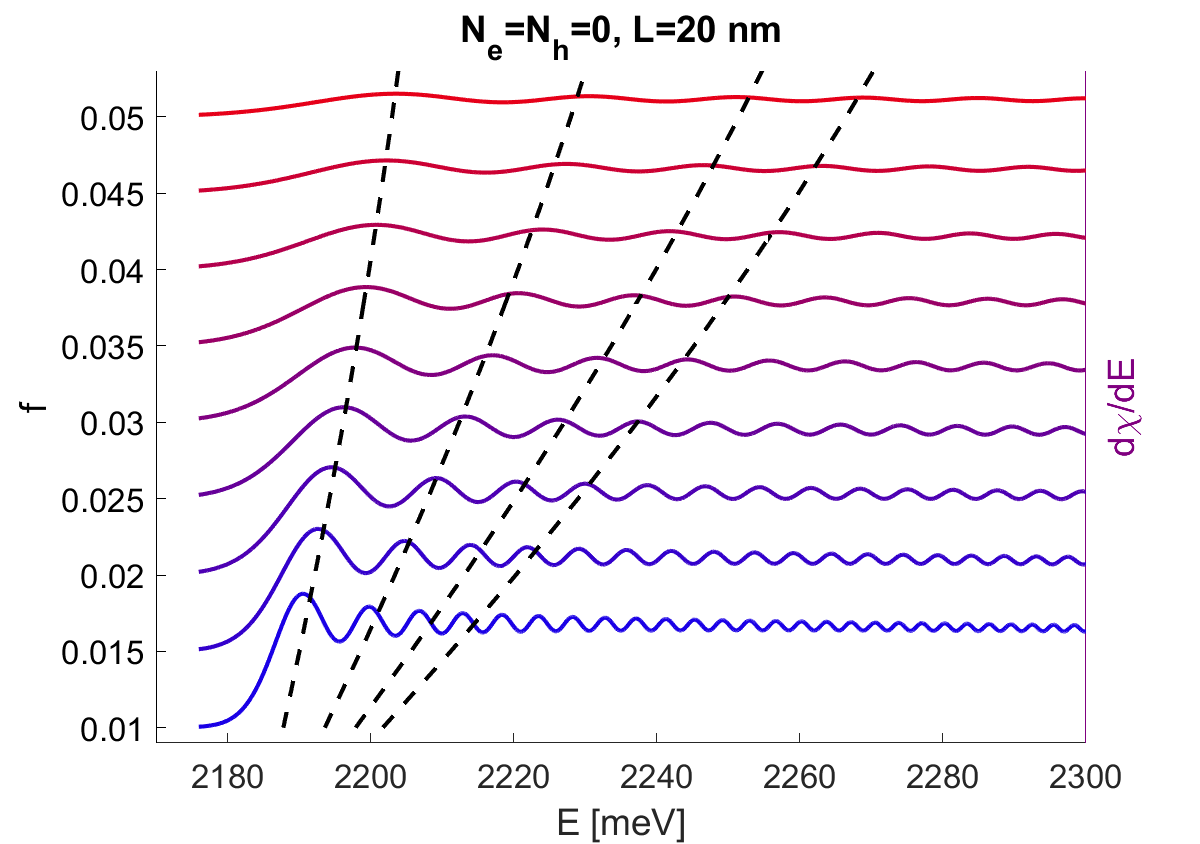}
b)\includegraphics[width=.8\linewidth]{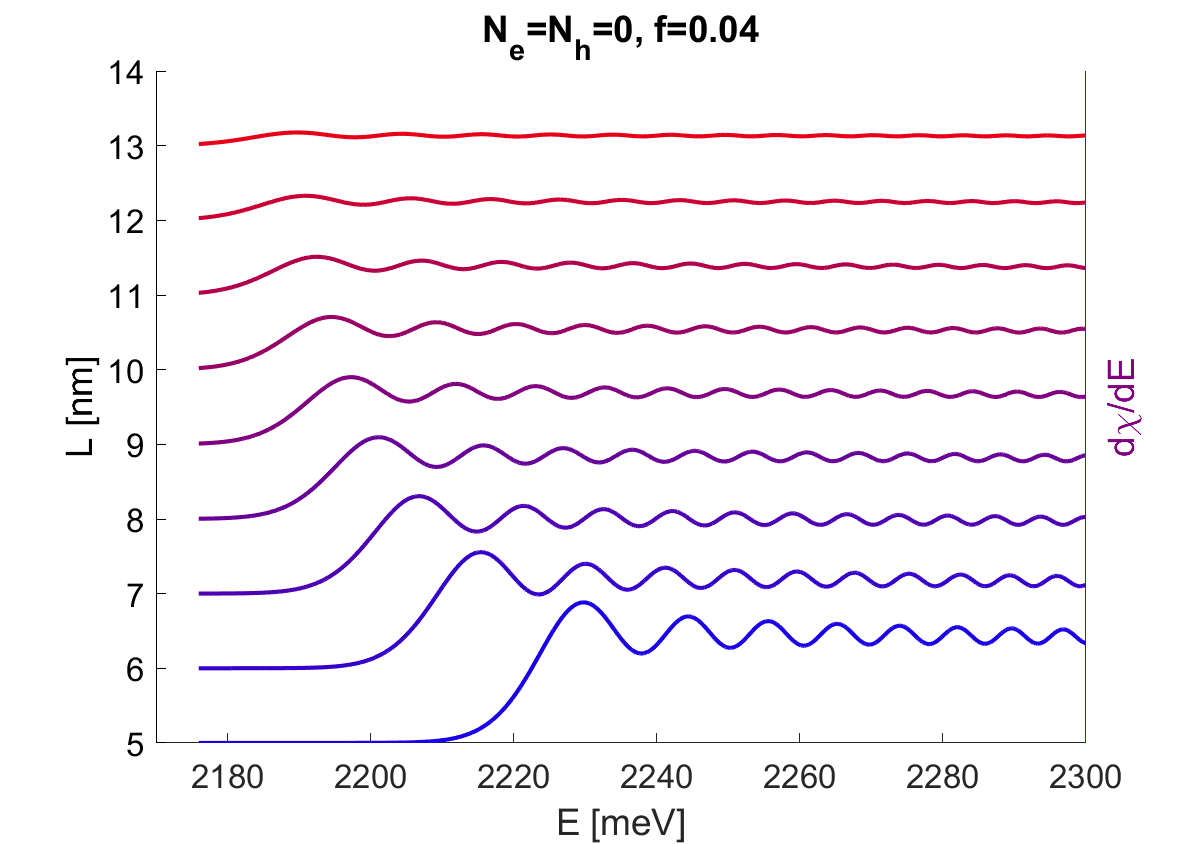}
c)\includegraphics[width=.8\linewidth]{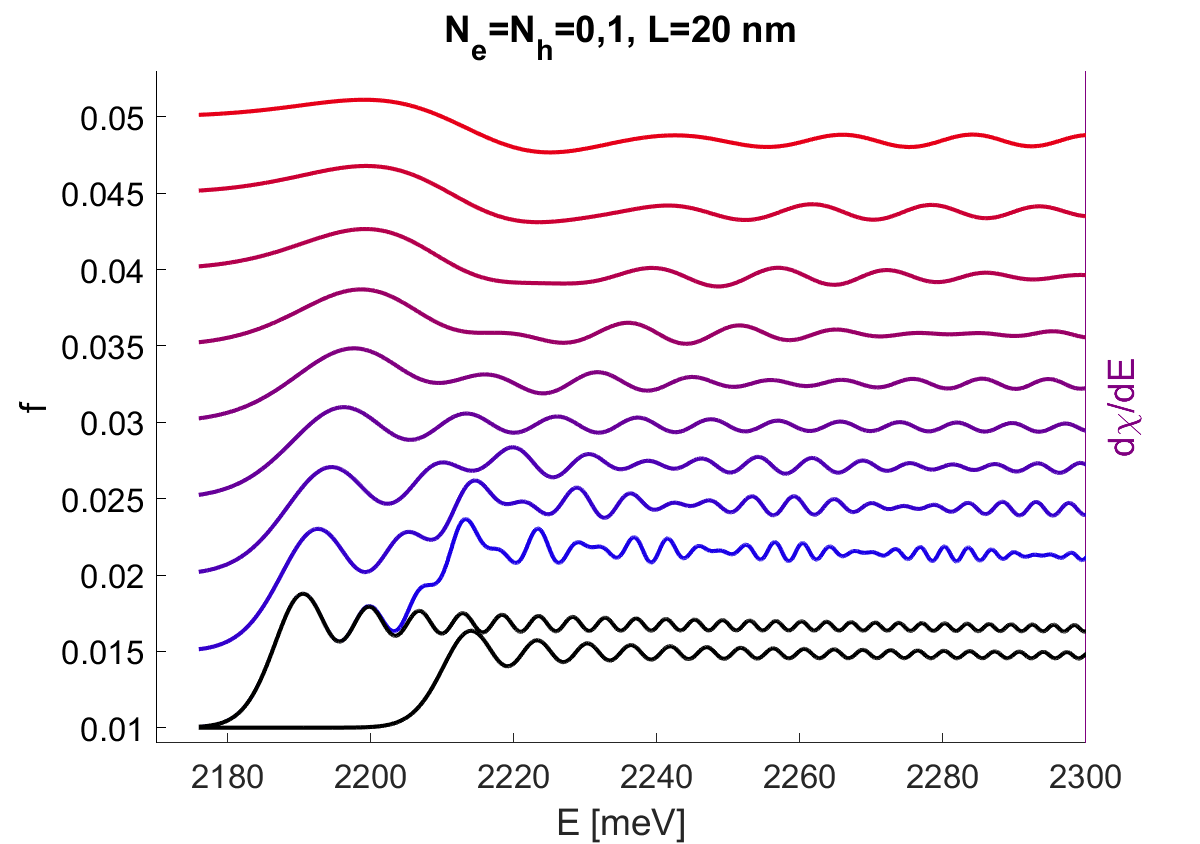}
d)\includegraphics[width=.8\linewidth]{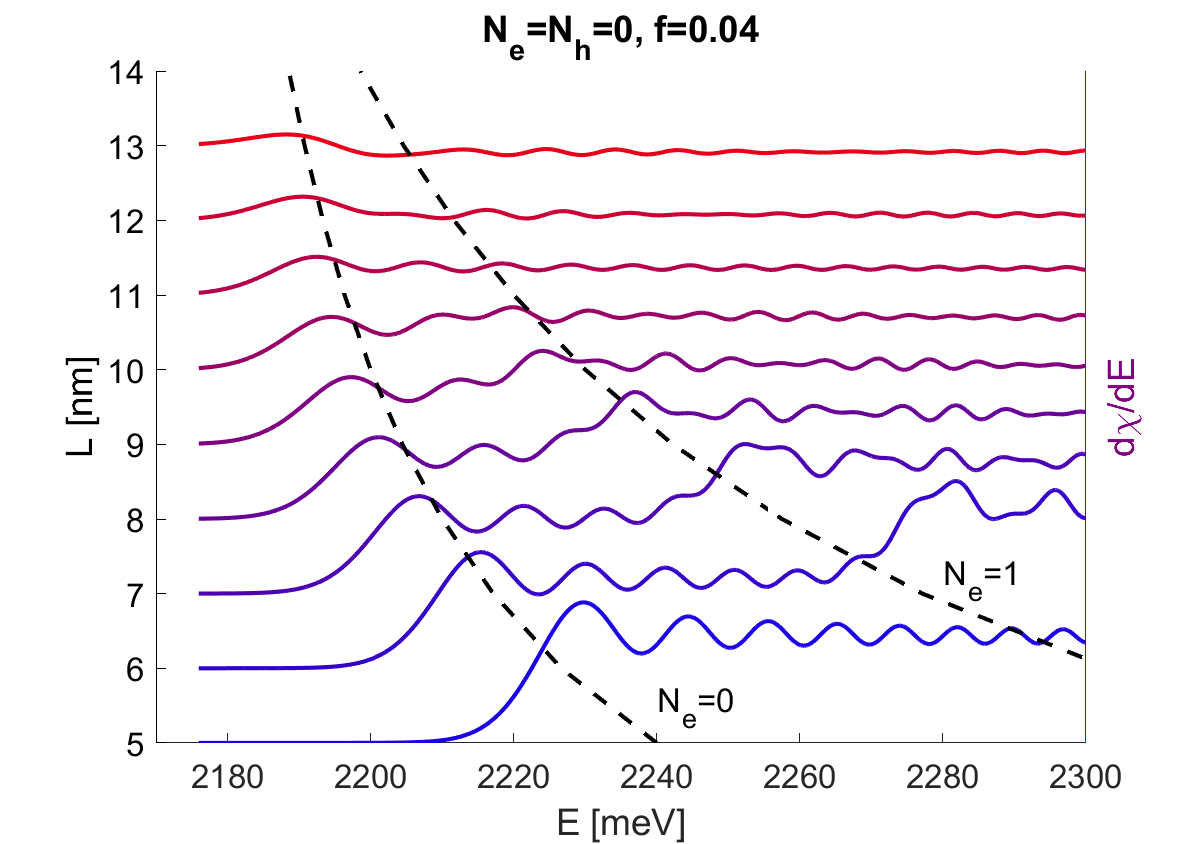}
\caption{Imaginary part of susceptibility in the energetic region above the band gap, calculated from Eq. (\ref{chi2D}) for $L=10$ nm, $j=0..9$ and a),b) $N_e=N_h=0$ b) c),d)$N_e=N_h=0,1$}\label{Fig:f1}
\end{figure}

We have performed calculations for a Cu$_2$O quantum well of
thickness $L=10 nm$, in the energetic region above the fundamental and
gap.  Fig. \ref{Fig:f1} presents the imaginary part of
susceptibility calculated from Eq. (\ref{chi2D})
in the region above the band gap; similarly to the bulk case
\cite{SZR FK}, the Franz-Keldysh oscillations occur.
Due to the low amplitudes of these oscillations and strong increase of overall absorption above the band gap, the first derivative of susceptibility is taken for better visibility. On  Fig. \ref{Fig:f1} a) one can see the oscillations of the absorption induced by various values of electric field f. The positions of the first four maxima are marked with black lines. One can see that the period of these oscillations decreases with $f$ and it is consistent with the results calculated for the bulk case \cite{SZR FK}. However, as compared to the bulk medium, the oscillation amplitude decreases with increasing $f$. In particular, the initial increase of absorption is much steeper for low $f$, similar to results obtained in \cite{Perebeinos}. The maxima are not evenly spaced; their energy is $E_{max} \sim n_{FK}^{2/3}$, where $n_{FK}$ is the number of maximum (Eq. (\ref{chi_okres})). Fig. \ref{Fig:f1} b) shows the dependence of the FK effect on the thickness $L$. One can observe an energy shift of the order of $\Delta E \sim 50$ meV for $L=5$ nm that is inversely proportional to $L$. The oscillation period is unaffected by $L$. On  Fig. \ref{Fig:f1} c) additional confinement state $N_e=N_h=1$ is included. The two black spectra at the bottom correspond to $N_e=0$ and $N_e=1$; the higher confinement is shifted by $\sim 20$ meV and both exhibit almost identical oscillations. Their sum creates a characteristic interference pattern, with a slow sine-like modulation caused by a small difference of the frequencies, similar to results presented in \cite{Xia2}. Finally, on  Fig. \ref{Fig:f1} d) one can observe that the higher confinement state reacts more strongly to the reduction of thickness $L$. In general, the system is capable of producing complicated interference patterns, with two degrees of freedom ($L$ and $f$) available for tuning them.

\begin{figure}[ht!]
\centering
a)\includegraphics[width=.8\linewidth]{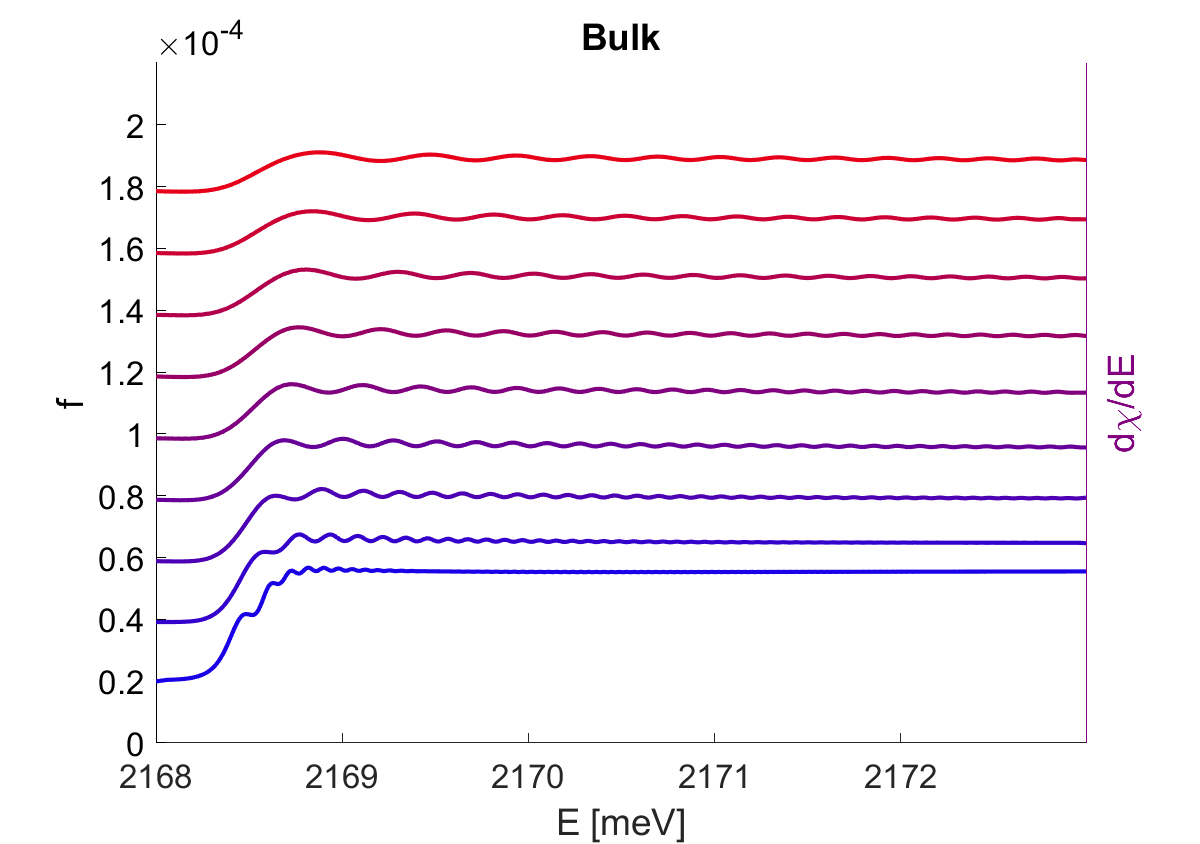}
b)\includegraphics[width=.8\linewidth]{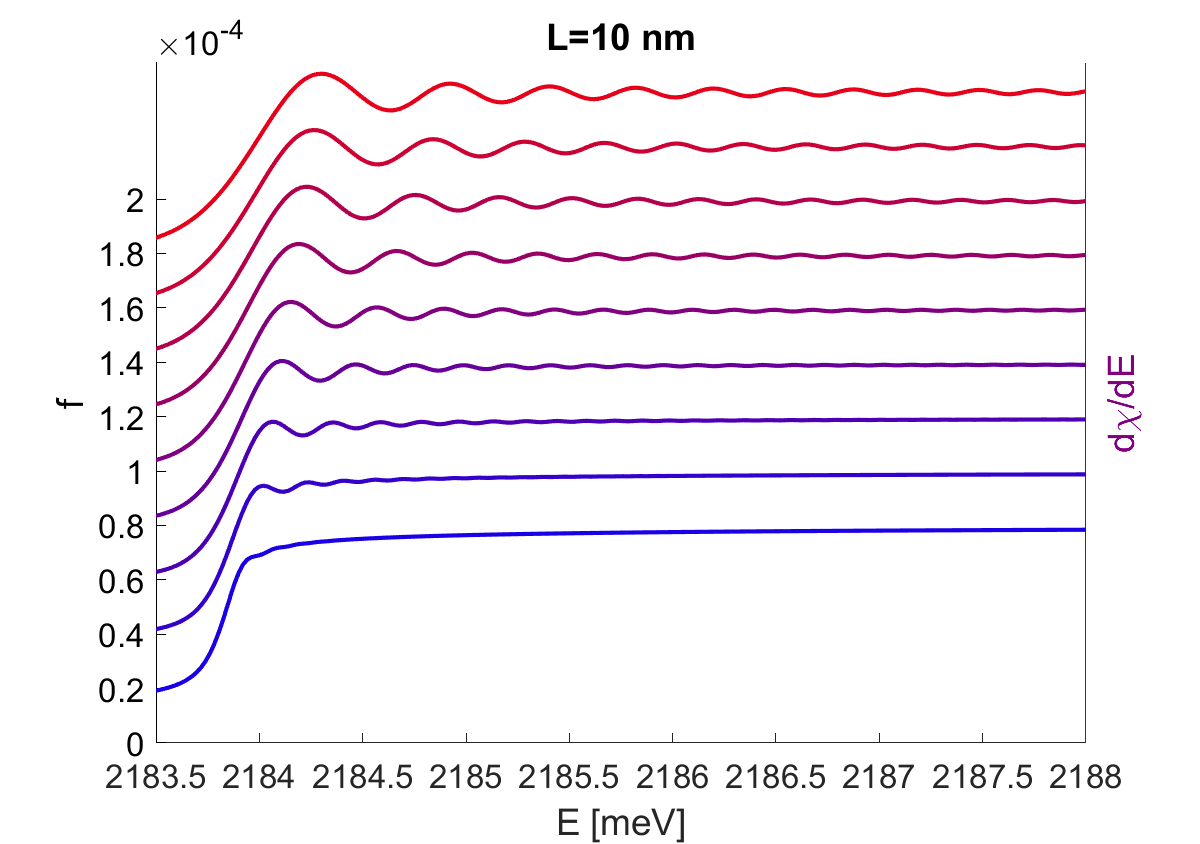}
c)\includegraphics[width=.8\linewidth]{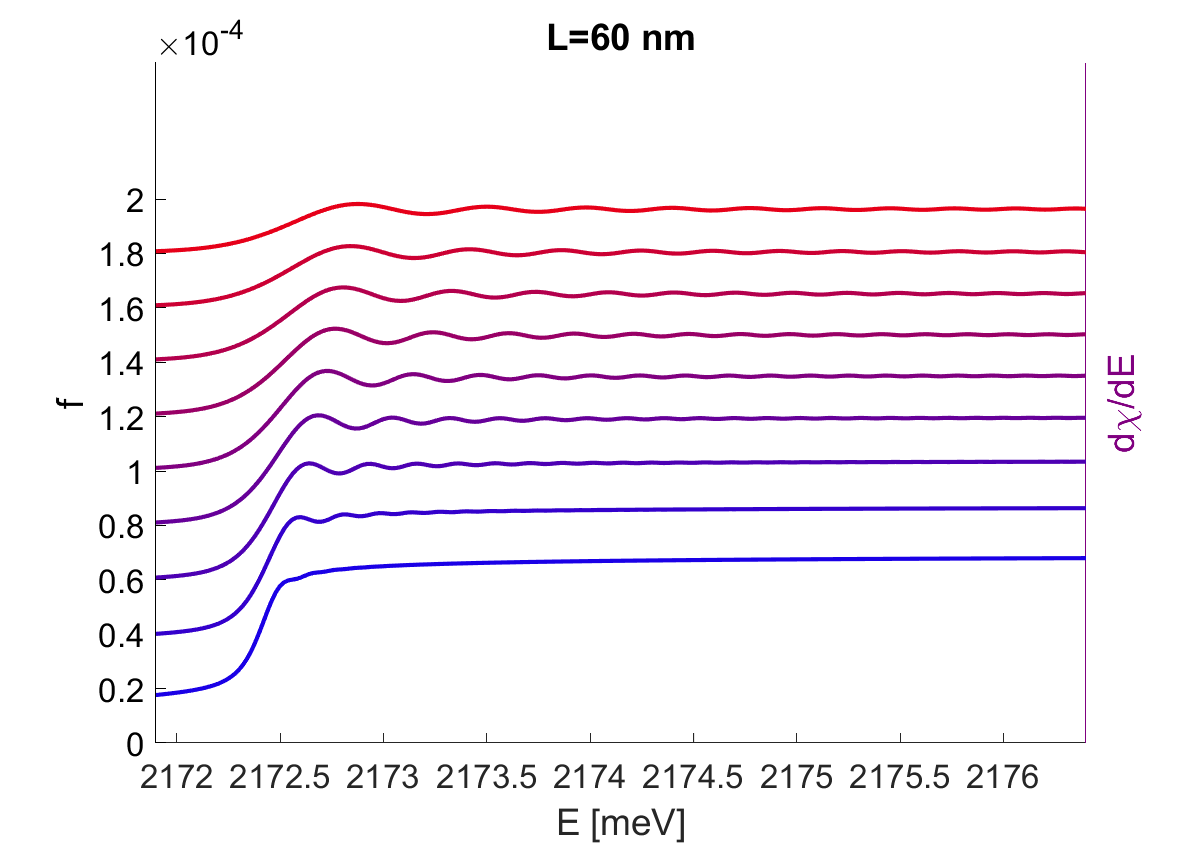}
\caption{Imaginary part of susceptibility in the energetic region above the band gap, comparison of a) bulk results and b) 10 nm and c) 60 nm well.}\label{Fig:f2}
\end{figure}

\begin{figure}[ht!]
\centering
a)\includegraphics[width=.8\linewidth]{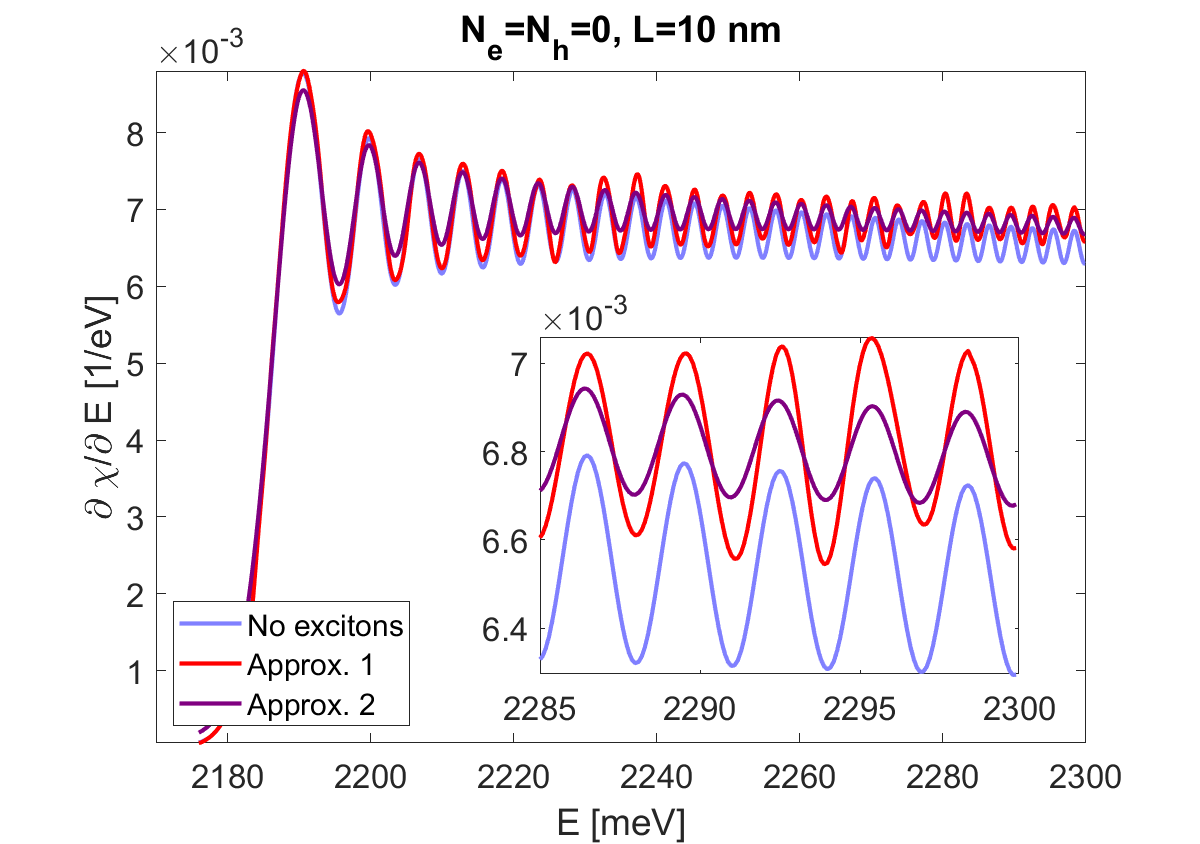}
b)\includegraphics[width=.8\linewidth]{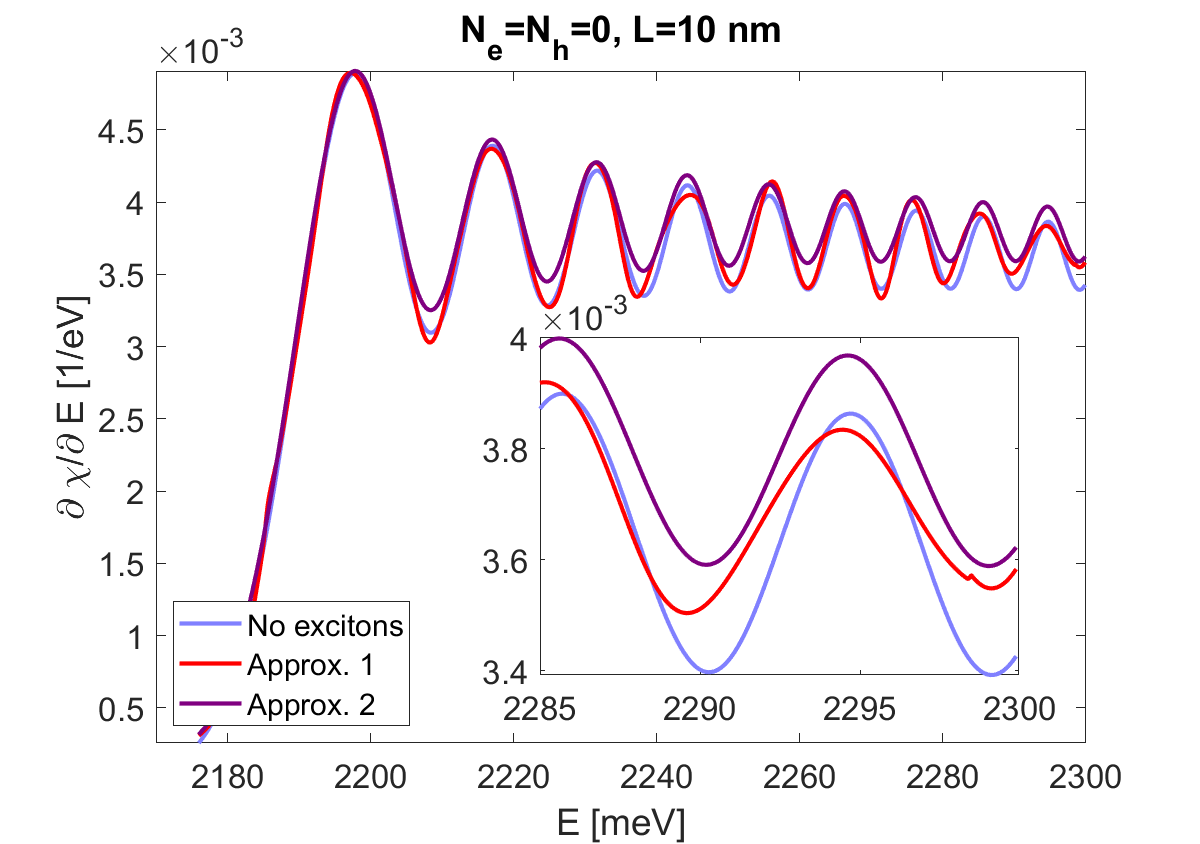}
\caption{Imaginary part of susceptibility in the energetic region above the band gap, calculated for a) $f=0.001$ and b) $f=0.005$, from Eq. (\ref{chi2D}) (no excitons), Eq. (\ref{chi2D_Q}) with Eq. (\ref{eq:denomin}) (Approx. 1) and Eq. (\ref{eq:denomin2}) (Approx. 2).}\label{Fig:f3}
\end{figure}

 Fig. \ref{Fig:f2} shows a comparison of the calculation
results with the spectra obtained in \cite{SZR FK}. Again, first
derivative of susceptibility is used for clarity. For the case of
low well thickness (Fig. \ref{Fig:f2} a)), one can notice several differences. In the quantum well, there is an increase of the
absorption slope ($\partial\chi/\partial E$) before the first
maximum of oscillation, which is less pronounced in the bulk results.
Moreover, the oscillations vanish at the limit of a very small
value of $f$ due to the fact that in the limit of $f=0$, the susceptibility becomes a slowly increasing function of $(E-E_g)$. For the stronger fields the oscillations are
more pronounced than in the bulk case. Due to the fact that $f<<1$, $f^{2/3}$>$f^1$ and the expression for the susceptibility in confined system in Eq. (\ref{chi2D}) has an amplitude that is overall higher and slower decreases with $f$. Notably, the period of oscillations is the same
in the bulk and QW, which is consistent with the estimation in Eq. (\ref{chi_okres}).
Due to the finite thickness the maxima in case of quantum well are shifted by a
constant energy $\Delta E \sim 20$ meV. By increasing L to 60 nm,
one obtains a result much closer to the bulk case (Fig.
\ref{Fig:f2} b)). In such a case, the amplitude of oscillations is
smaller in the QW due to the fact that $Im \chi \sim \frac{1}{L}$
Additionally, the increase of $\partial\chi/\partial E$ before the first maximum is much weaker
than for $L=10$ nm.
The addition of the resonant term $Q$ given by Eq. (\ref{eq:denomin}) into the expression for susceptibility (Eq. (\ref{chi2D_Q})) introduces a correction to the spectrum; due to the numerical difficulties connected with integration of functions containing potential singularities, one can also change the confinement potential to derive an alternative solution in Eq. (\ref{eq:denomin2}). Both solutions are shown on the Fig. \ref{Fig:f3} a). In general, the Coulomb interaction of electron-hole pairs
generates not only bound states below the gap, but also affects the continuum above the gap, with the effect of specific excitons proportional to their oscillator strength. One can see that inclusion of excitons does not affect the period of oscillations. Both approximations predict some reduction of oscillation amplitude (larger for modified potential) and a small overall increase of $\partial \chi/\partial E$ (roughly the same for both methods). The amplitude decreases monotonically for the case without excitons and approximation 2, while for the first approximation there are slight local variations (most noticeable for $E=2230$ meV and $E=2280$ meV). The lack of localized oscillatory terms in the second approximation is directly linked to the fact, that in approximation in Eq. (\ref{inny_pot}) there are no integrals over Airy functions.
In the stronger field regime (Fig. \ref{Fig:f3} b)) both approaches are in agreement regarding the amplitude. However, one can observe a slight phase shift for the results of Eq. (\ref{chi2D_Q}). The consistency of both approaches indicate that the underlying assumption of the second approximation, e. g. $f \to 0$, still holds for $f=0.005$. As mentioned before, the inclusion of excitons increases the mean value of $\chi$, especially in the high energy region (inset).

\section{Conclusions} 
The calculation results indicate that the general features of Franz–Keldysh effect are preserved in the QW system. Continuous, sinusoidal oscillations in the absorption spectrum emerge above the band gap, with a period dependent on the strength of applied electric field. This effect - a change in the absorption spectrum caused by an applied electric field - is the principle of operation of electro-modulator. By reducing the size of the system, one introduces multiple confinement states, which produce thickness-dependent interference effects in the F-K spectrum. Thus, QW system adds another degree of freedom in controlling the absorption spectrum. Finally, it is shown that the presence of excitons affects the F-K spectrum in a smooth, global manner, with no localized effects such as the excitonic lines below the gap. Overall, Rydberg excitons in Cu$_2$O enable convenient control and precise steering of absorption. The construction of high-sensitivity, compact  modulators based on quantum wells is the desirable task for modern quantum engineering.


\begin{thebibliography}{99}
\bibitem {Kazimierczuk}
T. Kazimierczuk, D. Fr\"{o}hlich, S. Scheel, H. Stolz, and M.
Bayer, Nature \textbf{514}, 344 (2014).

\bibitem{AssmannBayer_2020}
M. A{\ss}mann and M. Bayer, Semiconductor Rydberg Physics,
Advanced Quantum technologies, 1900134 (2020). DOI:
10.1002/qute.201900134.

\bibitem{Khazali}
M. Khazali, K. Heshami, and C. Simon, J. Phys. B \textbf{50}, 215301 (2017).

\bibitem{Walther}
V. Walther, R. Johne, and T. Pohl, Nat. Commun. \textbf{9}, 1309 (2018).

\bibitem{SZR FK}
S. Zieli\'{n}ska-Raczy\'{n}ska, D. Ziemkiewicz, and G. Czajkowski,
Phys. Rev. B \textbf{97}, 165205 (2018).

\bibitem{pss}
S. Zieli\'{n}ska-Raczy\'{n}ska, D. Ziemkiewicz, K. Karpiński and G. Czajkowski, Phys. Status Solidi B, 1800502, (2019).

\bibitem{Franz}
W. Franz, Z. Naturforsch A \textbf{13}, 484, (1958).

\bibitem{Keldysh}
V. Keldysh, Zh. Eksp. Teor. Fiz. \textbf{34}, 1138 (1958), Sov. Phys. JETP \textbf{7}, 788, (1958).

\bibitem{Cavalini}
A. Cavallini, L. Polenta, M. Rossi, T. Stoica, R. Calarco, R. Meijers, T. Richter,  H. L\"{u}th, Nano Lett. \textbf{7(7)}, 2166-2170 (2007).

\bibitem{Li}
D. Li, J. Zhang, Q. Zhang, Q. Xiong, Nano Lett. \textbf{12(6)}, 2993-2999 (2012).

\bibitem{Xia3}
C. Xia, S. Wei, H. N. Spector, Physica E: Low-dimensional Systems and Nanostructures \textbf{42(8)}, 2065-2068 (2010).

\bibitem{Perebeinos}
V. Perebeinos, P. Avouris, Nano Lett. \textbf{7(3)}, 609-613 (2007).

\bibitem{Miller}
D. A. B. Miller D. S. Chemla, and S. Schmitt-Rink, Phys. Rev. B \textbf{33}, 6976, (1986).

\bibitem{Xia}
C. Xia, and H. Spector,  Journal of Applied Physics \textbf{105}, 084313 (2009).

\bibitem{Xia2}
C. Xia, S. Wei, and H. Spector, Physica E \textbf{42}, 2065-2068 (2010).

\bibitem{Dow}
J. B. Dow, D. Redfield, Phys. Rev. B \textbf{1}, 3358 (1970).

\bibitem{DZ}
D. Ziemkiewicz, K. Karpiński, G. Czajkowski, and S. Zieli\'{n}ska-Raczy\'{n}ska,arXiv Electro-optical properties of excitons in Cu$_2$O
quantum wells: I discrete states, (2021).

\bibitem{Hamid}
K. Orfanakis, S. K. Rajendran, and Hamid Ohadi,arXiv:2011.12006v2
[cond-mat.mes-hall] 25 Nov 2020.

\bibitem{Abramowitz}
M. Abramowitz and I. Stegun, \emph{Handbook of Mathematical
Functions} Dover Publications, New York, 1965.

\bibitem{Ziemkiewicz_2020} D. Ziemkiewicz , K.
Karpi\'{n}ski, G. Czajkowski, and S. Zieli\'{n}ska-Raczy\'{n}ska,
Phys. Rev. B \textbf{101}, 205202, (2020).

\bibitem{maser}
D. Ziemkiewicz, S. Zielińska-Raczyńska, Optics Letters, \textbf{43}, 3742, (2018).

\end{thebibliography}
\end{document}